\documentclass[preprint,aps,prd,preprintnumbers,amsmath,amssymb,nofootinbib,superscriptaddress,floatfix]{revtex4-2}
\usepackage{framed}
\usepackage{bbold}
\usepackage{slashed}
\usepackage{graphicx}
\usepackage{amsmath}
\usepackage{amssymb}
\usepackage{epsfig}
\usepackage{hhline}
\usepackage{soul}
\usepackage{tabularx,pifont,multirow}
\usepackage{enumitem}
\usepackage{colordvi}
\usepackage{soul}
\usepackage{xcolor}
\usepackage{yfonts}

\newcommand{\be}{\begin{equation}}
\newcommand{\ee}{\end{equation}}
\newcommand{\cb}{\color{blue}}

\newcommand{\cy}{\color{white}}
\newcommand{\cbl}{\color{black}}

\usepackage[most]{tcolorbox}

\begin{document}

\title{How Much Can Gravitons Be Squeezed?}

\author{Panagiotis Dorlis} 
\email{panos$_$dorlis@mail.ntua.gr, psdorlis0@gmail.com} 
\affiliation{Physics Division, School of Applied Mathematics and Physical Sciences, National Technical University of Athens, Zografou Campus GR15780, Athens, Greece}

\author{Nick E. Mavromatos} 
\email{nikolaos.mavromatos@kcl.ac.uk} 
\thanks{Corresponding author}

\affiliation{Physics Division, School of Applied Mathematics and Physical Sciences, National Technical University of Athens, Zografou Campus GR15780, Athens, Greece}

\affiliation{Theoretical Particle Physics and Cosmology Group, Physics Department, King's College London, Strand, London WC2R 2LS, UK}

\affiliation{Associate member, Institute for Astronomy, Astrophysics, Space Applications and Remote Sensing, National Observatory of Athens, 15236 Penteli, Greece}

\author{Sarben Sarkar}
\email{sarben.sarkar@kcl.ac.uk} 

\affiliation{Theoretical Particle Physics and Cosmology Group, Physics Department, King's College London, Strand, London WC2R 2LS, UK}

\author{Sotirios-Neilos Vlachos} 
\email{sovlacho@gmail.com} 
\affiliation{Physics Division, School of Applied Mathematics and Physical Sciences, National Technical University of Athens, Zografou Campus GR15780, Athens, Greece}

\begin{abstract}
 Quantum Gravity remains elusive, largely because its observable effects are 
suppressed by powers of the Planck scale. Direct detection of single gravitons is 
widely believed to be impossible. Here we propose a concrete astrophysical mechanism 
that may overcome this suppression. We show that superradiant axion-like-particle 
clouds surrounding rotating black holes can generate multimode squeezed states of 
gravitons containing up to $10^6$–$10^7$ correlated quanta. Such states exhibit 
distinctive polarization correlations and quantum-noise signatures that could be 
detectable in future gravitational-wave interferometers. Observation of these 
signatures would constitute direct evidence for the quantum nature of gravitational 
radiation. Conversely, their absence can place constraints on axion-cloud lifetimes. 
Our approach also provides a test of General Relativity as an effective field theory.
\\
\begin{center}
{\rm Essay awarded Third Award in the Gravity Research Foundation 2026 Awards for Essays on Gravitation \\
(Submission date: March 21, 2026)}
\end{center}
\end{abstract}


\maketitle

\newpage
\pagestyle{plain} 

\subsection*{\emph{Circumventing the Suppression problem of Quantum-Gravity Phenomenology}}
\vspace{-0.4cm}

The central obstacle in quantum-gravity (QG) phenomenology is the \emph{suppression problem}. 
Any potentially observable quantum-gravity effect, at an energy scale $E$ for a low-energy observer,
is expected to scale as a power of the ratio 
$E/M_{\rm Pl}$, where $M_{\rm Pl}\simeq 2.435\times10^{18}\,{\rm GeV}$ is the reduced Planck mass, which is the characteristic scale where quantum gravity effects are expected to set in. This is quite a general remark for every Effective Field Theory (EFT) of QG, if the latter exists, either arising from an Ultra-Violet (UV) completion in the form of non-minimal couplings or from the nonlinearities of general relativity (GR) on its own~\cite{Donoghue:1994dn}.
For laboratory, astrophysical, or even gravitational-wave (GW) energies this ratio is extremely 
small, typically $E/M_{\rm Pl}\sim10^{-38}$ for, \emph{e.g.}, the energy scales of the LIGO experiment on the observation of GW from black-hole mergers~\cite{LIGOScientific:2016aoc}. As a result, even when GW are detected with exquisite precision, their (potential) quantum properties remain effectively invisible, unless appropriate physical circumstances or QG-models are discovered, which provide significant enhancement of the QG effects.

One such approach~\cite{Amelino-Camelia:1996bln,Amelino-Camelia:1997ieq,Ellis:1999rz,efmmn} entails Lorentz-Invariance-Violating (LIV) modified dispersion relations of ordinary radiation and matter probes,
due their propagation in quantum fluctuating space-time (QG ``foam''). In some circumstances, involving distant extreme-high energy cosmic probes, and/or intense astrophysical phenomena, such as Active Galactic Nuclei and Gamma Bursts, the 
sensitivity of some models has reached, or even surpassed, Planck energy scales. In this approach, the large cosmic distance and the high energy of the probes constitute the enhancement factors. Despite intense searches using  multi-messenger astrophysical and cosmological probes to date~\cite{Amelino-Camelia:1999hpv, Addazi:2021xuf}, no such phenomena have been discovered as yet. Nonetheless, the interest in probing LIV models of QG continues~\cite{AlvesBatista:2023wqm}.

If gravity is considered as a  fundamental quantum force, which is our point of view, then its most direct experimental evidence would be the detection of single gravitons, the quantum field carrier of the gravitational interaction, in analogy with the case of quantum electrodynamics, whose carrier, the photon, has long been experimentally detected and most of its properties understood. In fact, current quantum-optics technology allows the manipulation of single quantum photons, which can then find important applications to quantum information theory, quantum computing, and precision detection.

However, unlike photons, the detection of single gravitons is extremely difficult, and according to Dyson~\cite{Dyson:2013hbl}, probably impossible, to achieve. Dyson has argued that such a detection would require Planck-scale sensitivity of a strain detector, like the LIGO interferometer~\cite{LIGOScientific:2016aoc}, which would lead to a catastrophic gravitational collapse, due to the formation of a black hole as a result of such a precision measurement. Essentially, Dyson argued that a single graviton absorption cross section is of order of the Planck area  $4\pi^2 \ell_P^2 $ (where $\ell_p = 10^{-33}$~cm, the Planck length), and therefore require sufficiently dense and massive detectors, with absorption cross section of order $10^{-41}~{\rm cm}^2$
per gram, which could create the conditions of gravitational collapse. 

However, there are states of the gravitational field, which constitute the gravitational analogues of 
the so-called \emph{squeezed states} of photons in quantum optics~\cite{WallsMilburn2008, Agarwal_2012}, which might contain a large number of gravitons, with no classical analog on their expected observables. In quantum optics, the average number of photons with respect to the ``squeezed vacuum'', obtained from the ordinary vacuum by an appropriate Bogoliubov  transformation (``squeezing operator"~\cite{WallsMilburn2008}), is $\langle n_{\rm ph}\rangle \sim {\rm sinh}^2(r)$, 
where $r$ is the so-called squeezing parameter, which might be a large number under appropriate circumstances. 
In \cite{Parikh:2020nrd,Parikh:2020kfh,Parikh:2020fhy,Abrahao:2023lle} it has been suggested that, if quantum gravitons are produced as single-mode squeezed states in GW, their detection might not be impossible in interferometers, through measurements of their \emph{stochastic noise}~\cite{Amelino-Camelia:2001dbf}. In situations where the corresponding squeezing parameter $r$, would be relatively large, therefore, one might be able to overcome Dyson's single-graviton obstruction. 

However, until now, concrete sources of squeezed states of gravitons have not been identified, except for the case of cosmic inflation~\cite{Grishchuk:1989ss,Grishchuk:1990bj,Albrecht:1992kf,Mukhanov:2007zz}, where the amplification of quantum fluctuations into macroscopic cosmological perturbations (primordial inhomogeneities in the energy density of the early Universe) is a process of quantum-squeezing creation. Its efficient detection though is still not feasible. In general, squeezed graviton states are expected to be suppressed, characterised by a squeezing parameter $r \sim (E/M_{\rm Pl})^n$, $n \in \mathbb Z^+$. Moreover, as discussed in \cite{Carney:2023nzz,Carney:2024dsj}, as far as the case of GW is concerned, the effects of the aforementioned single-mode  squeezed quantum-graviton states on the interferometer might not be distinguishable from those of a classical GW.

This Essay proposes a mechanism capable of overcoming the suppression 
effects of QG by identifying a concrete astrophysical source of strongly 
squeezed gravitons. We show that superradiant axion clouds surrounding 
rotating black holes can generate multimode squeezed graviton states 
with an effective squeezing parameter reaching $r\sim60$--$70$, 
corresponding to $10^{6}$--$10^{7}$ correlated gravitons. Such large 
collective quantum states dramatically amplify otherwise unobservable 
quantum-gravity effects and may produce detectable signatures in 
future gravitational-wave interferometers (see Fig.~1 and titled paragraph
 below). The exploration of astrophysical sources of squeezed 
gravitons has only recently begun to appear in the literature 
\cite{Kanno:2025how,Manikandan:2025dea,Guerreiro:2025mcu}. 

\begin{figure}[t]
    \centering
\includegraphics[width=0.6\linewidth]{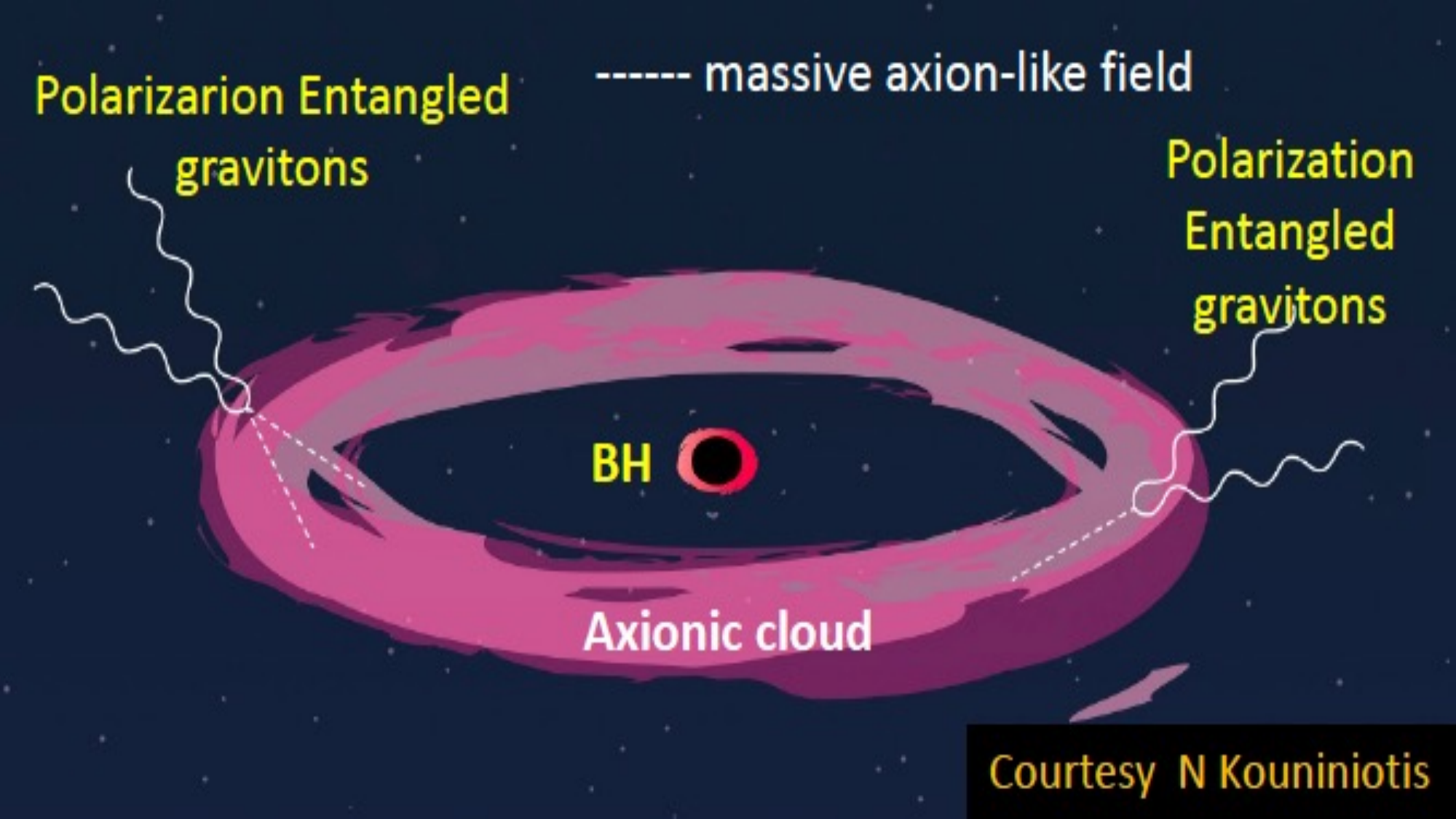}
    \caption{A rotating (astrophysical) black hole (central blob (BH)) and its surrounding axion-cloud condensate (lightly coloured ring). The two processes of the production of pairs of squeezed entangled graviton states (wavy lines) are indicated: fusion of two axions (dashed lines), which is the dominant General-Relativity effect, and a decay of an axion, which is the result of the (subdominant) Chern-Simons anomalous term.}
    \label{cloud}
\end{figure}

\begin{tcolorbox}[
colback=white,
colframe=black,
title=\textbf{\cy The Squeezed Graviton Hypothesis \cb },
fonttitle=\bfseries,
boxrule=0.8pt,
arc=2mm,
left=2mm,
right=2mm,
top=1mm,
bottom=1mm
]

Rotating black holes, surrounded by superradiant axion clouds, can act as astrophysical nonlinear media that generate collective multimode squeezed states of 
gravitons~\cite{Dorlissuperr,Dorlis:2025amf} (see fig.~\ref{cloud}). Indeed, in such systems (which are characterized by a macroscopic occupation number of axions that can survive for a long time) the gravitational field can enter a highly squeezed quantum state containing a macroscopic number of correlated gravitons. This dramatically amplifies otherwise unobservable quantum-gravity effects.
The key observational consequence of this hypothesis is that the emitted 
gravitational radiation is not purely classical but contains distinctive 
quantum signatures. In particular, gravitons are predicted to be 
entangled with characteristic polarization correlations and modified 
quantum-noise properties. Detection of such squeezed graviton states in 
future interferometers would therefore provide direct evidence for the 
quantum nature of the gravitational field.

\end{tcolorbox}

 \subsection*{\emph{Squeezing is everywhere: From Crystals and Lasers to the Gravitational Atom}}\label{sec:squeez}
\vspace{-0.4cm}

 From various observations and theoretical simulations,  GW detected in 2015 have been interpreted as arising from merging celestial objects consistent with rotating black holes (BH). 
A culmination of the study of BH was the photograph of the black hole at the centre of the nearby Galaxy M87* by the Event Horizon Telescope (EHT) collaboration~\cite{EHT1,EHT2}. BH, due to their strong gravity regimes, are believed by many to be the best laboratories for testing QG models. This will be our basic point of view in our quest for detecting QG. 
Specifically, we  exploit 
the mechanism presented in~\cite{Dorlissuperr,Dorlis:2025amf}, in which \emph{multi-mode} squeezed graviton states can be produced in an axion-cloud environment that can potentially surround astrophysical rotating black holes (see fig.~\ref{cloud}, which illustrates the qualitative geometry of the axion cloud condensate). Under such circumstances, the axions form a condensate-like distribution at the exterior of a rotating black hole, which develops a \emph{superradiance} instability~\cite{Detweiller,BritoCardoso,BritoScales1}. This leads to a large occupation number of axions in a long lived quasi-stationary phase, which are the basic features for a potentially large squeezing.

In quantum optics~\cite{WallsMilburn2008, Agarwal_2012} squeezed photons are generated through nonlinear 
interactions of the electromagnetic field in a medium such as nonlinear 
crystals, optical fibers, or atomic vapours. In spontaneous parametric 
down conversion (SPDC) ~\cite{PhysRevA.31.2409, Wu:1986zz} a pump photon splits into two entangled photons, 
while in spontaneous four-wave mixing (SFWM) two pump photons convert 
into an entangled pair. The \emph{large occupation number} of the pump field 
enhances the squeezing process.
Squeezing is produced because the created photon pairs are quantum correlated, with the relevant quantum correlations manifesting themselves as squeezing in the electric field quadratures.
In both processes, the large occupation number of photons in the coherent state  increases  squeezing, with SFWM having a larger squeezing, due to its quadratic dependence on the amplitude of the field, as compared to the linear dependence in SPDC.  

In the framework of weak QG EFT~\cite{Donoghue:1994dn}, GR induces an infinite number of higher order self-interactions for the graviton. This is also the case, when one considers minimal or non-minimal couplings with matter fields. These interactions can  lead to squeezed gravitons in the non-relativistic regime and with energy-momentum conservation at the vertex, 
in full analogy with the SPDC and SFWM processes in quantum optics. In summary, 
the axion condensate plays the role of the pump field in quantum optics, while the nonlinearities induced by crystals are an analogue of 
the gravitational interaction.

The rotating BH superradiance is  
a \emph{runaway superradiance}~\cite{Detweiller,BritoCardoso,BritoScales1}, characterised by an instability; particles and electromagnetic radiation scattering from a rotating (Kerr-type) BH,
can gain energy and angular momentum in the process. If this radiation is reflected back at the black hole,  a runaway explosive process 
(sometimes called a “black-hole bomb”)  
could develop.  This requires the presence of a massive (pseudo) scalar (axion) field. 
In a rotating (Kerr) BH background, this field exhibits 
an exponentially growing-mode \emph{instability}. The system of the axion field in the Kerr background spacetime has many similarities with the quantum system of the hydrogen atom, hence its name \emph{gravitational atom}. In this terminology, viable in the \textit{non-relativistic regime}, i.e. the Compton wavelength of the axion being much larger than the black hole outer horizon radius, the fastest growing mode corresponds to the $``2p"$ state of the hydrogen atom.

The condition for \emph{black hole superradiance}~\cite{Detweiller,BritoCardoso} reads,
\begin{align}\label{gravsrcond}
{\rm Re}(\omega_{n \ell m}) < m\, 
\Omega_H\,, 
\end{align}
where $(n, \ell, m)$ are the ``quantum  numbers of the gravitational atom'',  $\omega_{n \ell m}$ is the frequency of the axion field, 
$\Omega_H = \frac{\alpha}{\alpha^2 + r_+^2}$ is the angular velocity of the outer horizon of the black hole, $r_+$, and $\alpha = \mathcal J_{\rm H}/\mathcal M$, is
its angular momentum in units of the black-hole mass $\mathcal M$. 
 This gravitational instability is \emph{exponential} in nature, and is manifested in the existence of a \emph{positive} imaginary part of the axion-field frequency,
 ${\rm Im}(\omega_{\rm n \ell m}) =1/ \tau_s > 0$. In this sense, $\tau_s$ denotes the characteristic time scale during which superradiance is effective.

Upon extracting significant amounts of energy 
from the spinning black hole, the axions form 
 clouds of extremely high occupation number and finite, but very long life time $T\gg\tau_s$. The clouds are a kind of a
``condensate’’ in the exterior region of the outer horizon,  
extending to macroscopically large distances from the horizon.
Under such circumstances, there can be significant enhancement of the number of the produced squeezed-gravitons~\cite{Dorlissuperr,Dorlis:2025amf}, which, together with the \emph{entangled} (non-separable) nature of the \emph{multimode} gravitons, should constitute promising candidates for probing the quantum nature of the emitted GW. On the other hand, the non-observation of squeezed graviton states in current interferometers can place upper bounds on the lifetimes of the axionic clouds. Hence, this approach can also contribute to axion phenomenology.

Our starting point~\cite{Dorlissuperr,Dorlis:2025amf} is that, in the exterior neighborhood of a rotating (Kerr-type) astrophysical black hole, 
the gravitational EFT is described by the so-called Chern-Simons (CS) gravity~\cite{Jackiw:2003pm} action:\footnote{Our conventions are those of \cite{Dorlis:2025amf}.} 
\begin{align}\label{CSgravaction}
S = \frac{1}{2\kappa^2} \int d^4x \, \sqrt{-g}\, \Big[R  - \frac{1}{2} \partial_\mu\, b\, \partial^\mu b - \frac{1}{2} \mu_b^2 \, b^2  \Big]
-\frac{A}{2}\,\int d^{4}x\sqrt{-g}\ \ b \, R_{\mu\nu\rho\sigma}\widetilde{R}^{\nu\mu\rho\sigma} \,
+ \dots \,,   
\end{align}
where  $\kappa = M_{\rm Pl}^{-1}$ is the (3+1)-dimensional gravitational constant. $R$ denotes the  nontrivial spacetime curvature scalar, 
$\widetilde{R}_{\mu\nu\rho\sigma}=\frac{1}{2}R_{\mu\nu\alpha\beta}\ \epsilon^{\alpha\beta}\!_{\rho\sigma}\ $ is the dual of the Riemann tensor $R_{\mu\nu\alpha\beta}$, with  $\epsilon_{\mu\nu\rho\sigma}$ the covariant Levi-Civita tensor, and the field $b(x)$ corresponds to an axion-like-particle (ALP) of mass $\mu_b$. The massive nature of the axions is necessary for superradiance to occur~\cite{Detweiller,BritoCardoso}. 
The last term of the integrand of \eqref{CSgravaction} 
is the Chern-Simons (CS) gravitational anomaly, which is nontrivial in a rotating black-hole background, acting as a source of axions. 
The $\dots$ denote possible interactions of the ALP field, as well as other matter (or radiation) fields, which are not of direct relevance to us here. The coefficient of the CS term $A$ is inversely proportional to the axion coupling constant $f_b$, and thus depend on the kind of axion. For instance, in the context of string theory~\cite{str1,pol1,pol2}, the axions arising from compactification~\cite{Svrcek:2006yi}, depend on the specific string model underlying the EFT \eqref{CSgravaction}; an axion can also be string-model independent, corresponding to the gravitational bosonic multiplet of the string, acquiring a mass through some instanton effects of an appropriate non-Abelian gauge group. In the latter case, the coefficient is given by~\cite{Duncan:1992vz}:
\begin{equation}
\label{Adef}
A= 
  \sqrt{\frac{2}{3}}\frac{M_{\rm Pl}}{48\, M_s^2} \,,
\end{equation}where $M_s$ is the string mass scale~\cite{str1}. \\
The presence of a macroscopic number of ALPs in the cloud, during the lifetime of the cloud, is the main reason that such situations have an enhanced production of squeezed gravitons.
Since we work on the \textit{non-relativistic regime}, $G\mathcal{M}\mu_b\ll 1$, the cloud is concentrated far away from the event horizon. Thus, we shall adopt a weak graviton EFT about a flat Minkowski background, $\eta_{\mu\nu}$, $\mu, \nu =0, \dots 3$, based on the graviton field expansion: 
$g_{\mu\nu} = \eta_{\mu\nu} + \kappa h_{\mu\nu}$, where  $\kappa \vert h_{\mu\nu} \vert \ll 1$, is the metric-tensor (GW type) perturbation in the transverse and traceless (TT) gauge~\cite{tHooft:1974toh}. 
Adopting a \emph{canonical quantization} procedure for these perturbations,
and restricting ourselves to quadratic order in the weak-graviton $h_{\mu\nu}$ expansion, we identify the following leading interaction terms of the quantum graviton excitation fields $h_{\mu\nu}$ with the axions $b(x)$ in the effective action \eqref{CSgravaction}:
\begin{align}\label{S12}
    S^{(1)}&= \frac{\kappa}{2}\int d^4x \ h_{ij}T^{ij}  \ ,  
    \nonumber \\ 
    S^{(2)}&=  - \frac{\kappa^2}{2} \int d^4x \ h_{im}h^{m}_{\ j} \ \partial^i b\,\partial^j b \ ,  \nonumber \\
     S^{(2)}_{CS} & = -A\kappa^2\int d^4x \ b(x) \, \epsilon_{ijk} \ \Bigg( \partial_l\partial^k  h^j_m  \partial^m\dot{h}^{li}  +\ddot{h}^{li}\partial^k\dot{h}^j_l  
    -\partial_m\partial^kh^j_l\partial^m\dot{h}^{li}  \Bigg)   \,, 
\end{align}
where $T_{ij}$ is the matter $b$-axion field energy-momentum tensor, and the indices $i,j,k=1,2,3$ are spatial ones. The term $S^{(1)}$ produces coherent graviton states, which are close to being classical states, and therefore will have very suppressed quantum noise. This is also the same for the vacuum state. Therefore it is not of relevance to our purposes here. 
The interesting terms are the $S^{(2)}$ which produce entangled (non-separable) graviton states within the GR framework, through the fusion of two ALPs, and the $S^{(2)}_{CS}$ trilinear term, describing the production of an entangled graviton state from the decay of an ALP (see fig.~\ref{cloud})~\cite{Dorlissuperr,Dorlis:2025amf}. 
The latter term, arising from the CS gravitational anomaly, is beyond the GR framework.
The term $S^{(2)}$ 
is the gravitational \emph{analogue} of
the SFWM in quantum optics.  
On the other hand, the CS anomalous term $S^{(2)}_{CS}$ corresponds to the SPDC.

\vspace{-0.74cm}
\subsection*{\emph{Number of Squeezed Gravitons and Phenomenology}} \label{sec:numnbersqueezing}
\vspace{-0.4cm}

In our case it is squeezed gravitational radiation that is produced. 
The so produced gravitons are multimode (peaking as an approximate two-mode in this particular case where the non-relativistic regime is assumed) entangled (non-separable) quantum states~\cite{Dorlissuperr,Dorlis:2025amf}. 
The relevant evolution operator for our gravitational system pertaining to the GR contributions ({\emph cf.} $S^{(2)}$ term in \eqref{S12}) has the form of a squeezing operator~\cite{Scully_Zubairy_1997,WallsMilburn2008,Agarwal_2012}
\begin{equation}
\label{Scattering_Multimode_Squeeze_GR}
    \hat{S}^{(2)}_{GR}=\exp\left[ \frac{1}{2}\sum_{I,J}\mathcal{G}^{(GR)}_{IJ}\ \hat{\alpha}^\dagger_I\hat{\alpha}^\dagger_J -{\rm hermitian~conjugate} \right]  \, ,
\end{equation}
where $\hat{\alpha}_I^\dagger$ is the graviton creation operator, and the indices $I, J =(\lambda,\vec{k})$ denote the graviton states of helicity and three-momentum, respectively. 
For the rotating-black-hole-ALP system, this is the multimode squeezing operator familiar from quantum optics. The coefficient $\mathcal{G}^{(GR)}_{IJ}$ plays the role of a mode-dependent squeezing amplitude and is proportional to the time integral of the interaction Hamiltonian. Thus, for sufficiently long interaction time, where energy conservation is a valid approximation, the mode-dependent squeezing amplitude reads,
\begin{align}\label{GRsqop}
\mathcal{G}^{(GR)}_{IJ}\sim -2\,i\, \kappa^{2} \, N_{b} \, \mathcal{F}^{(GR)}_{IJ}\, T \,,
\end{align}
where $N_b$ is the axion occupation number in the cloud and $T$ its lifetime, i.e. the period of its quasi-stationary phase, while diluting through classical (coherent) GWs and $ \mathcal{F}^{(GR)}_{IJ}$ denotes the mode correlation function.
Because the axion cloud contains a macroscopic number of particles produced through superradiance, the factor $N_{b}$ can be extremely large. Consequently the effective multimode squeezing parameter $r$ of the gravitational radiation field has been estimated to lie in the range~\cite{Dorlissuperr,Dorlis:2025amf}
\begin{align}\label{GRsqr} 
\mathcal O(0.1) \lesssim  r^{2}\equiv \sum_{IJ} \left| \mathcal{G}^{(GR)}_{IJ} \right|^{2} \lesssim 2.5\times10^{-15} \,T \mu_b\,.
\end{align}
For sufficiently large $\mu_b T \sim {\mathcal O}\left(10^{16}\right)$~\cite{porto_scales}, 
$r$ can reach values up to $r\sim O\left( 60-70 \right)$, implying that the axion-cloud system can  generate gravitational radiation 
containing up to $10^6$–$10^7$ correlated gravitons, many orders of 
magnitude larger than estimates based on single-graviton processes. For the $2p$ state, $\mathcal F^{(GR)}_{IJ}$ has been calculated in \cite{Dorlis:2025amf}. 
The axionic condensate (cloud) around the Kerr (spinning) black hole, plays the r\^ole of the classical (coherent) source driving the process. The detailed analysis of \cite{Dorlis:2025amf} demonstrated that the 
dominant contributions to $\mathcal G_{IJ}^{(GR)}$ come from modes with $k \simeq k^\prime \simeq \mu_b$, and the 
generated entangled graviton pairs acquire ($L,R$) polarization entanglement, 
with the opposite polarization correlations being enhanced. 
This implies that the cross-polarization correlations of the entangled graviton states (of Einstein-Podolsky-Rosen (EPR)~\cite{epr} type) are approximately symmetric under the interchange of $L \leftrightarrow R$, i.e. $\vert \Psi_{GR}\rangle \approx \frac{1}{2}\, \mathcal{G}^{(GR)}_{(R,\vec{k})(L,\vec{k}^\prime)} \left(\vert RL\rangle+ \vert LR\rangle \right)$.

The CS anomaly-induced squeezing, associated with the CS anomalous term $S^{(2)}_{CS}$ in \eqref{S12}, can be studied in a similar manner~\cite{Dorlissuperr,Dorlis:2025amf}. Assuming the dominant contribution $k\approx k^\prime \approx\mu_b/2$, 
as in the case of GR, we obtain a maximal polarization-entanglement, in which only opposite helicity states are allowed to be produced in an antisymmetric Bell-state configuration, $
    \vert \Psi_{CS}\rangle = \frac{1}{2} \, \mathcal{G}^{(CS)}_{(R,\vec{k})(L,\vec{k}^\prime)}\left( \vert LR\rangle-\vert RL\rangle     \right)$, 
since $\mathcal{G}^{(CS)}_{(R,\vec{k})(L,\vec{k}^\prime)} = - \mathcal{G}^{(CS)}_{(L,\vec{k})(R,\vec{k^\prime})}$, as expected from the fact that the gCS anomaly terms in \eqref{CSgravaction} vanish identically for non-chiral GW. This case dominates when the emitted pairs have opposite projections on the BH and cloud rotation axis and a nearly anti-collinear emission.
If, on the other hand,  both gravitons have the same projection, then only pairs of the same helicity are allowed:
$\vert\Psi_{CS}\rangle=\frac{1}{2}\mathcal{G}^{(CS)}_{(L,\vec{k})(L,\vec{k}^\prime)}(\vert LL\rangle-\vert RR\rangle)$. 

If one assumes axions with anomalous couplings \eqref{Adef}, then it is easy to demonstrate that such terms are subdominant compared to the GR terms by several orders of magnitude, for all realistic values of the string mass scale $M_s$, black-hole mass and axion masses $\mu_b$~\cite{Dorlis:2025amf}. This was to be expected because the CS term is a higher derivative term, as compared with the GR terms, and moreover, it involves one ALP decay into two entangled graviton states, 
in contrast to the GR term, which involves the fusion of two ALPs into a pair of squeezed entangled gravitons. Hence, the CS enhancement due to the macroscopic number of $2p$ ALP states will be proportional to $\sqrt{N_b}$~\cite{Dorlis:2024yqw}, as compared with the $N_{b}$ factor of the GR~terms~$S^{(2)}$. 

It is important to stress here that the 
observational prospects in the squeezed graviton case at hand are significantly enhanced, as compared with searches 
of single graviton states~\cite{singlegrav,Carney:2023nzz} or single-mode squeezed states~\cite{Parikh:2020nrd,Parikh:2020kfh,Parikh:2020fhy,Manikandan:2025hlz,Manikandan:2025dea}. This is a consequence of the existence of a macroscopic number of axions in the cloud, due to its condensate nature, and its long lifetime. 
Moreover the entangled nature of the produced gravitons is unambiguously quantum in nature, once they are detected. We also note that in the relativistic case, where curvature effects cannot be avoided, the almost anti-collinear emission of graviton polarisation states breaks down. This may lead to additional enhancement of squeezing. 

\vspace{-0.74cm}

\subsection*{\emph{Comments on detection}}

\vspace{-0.4cm}

 We next remark that the non-observation of squeezed single-mode graviton states by LIGO/Virgo interferometers~\cite{LIGOScientific:2016aoc,McCuller:2021mbn} implies~\cite{Hertzberg:2021rbl} an upper bound on the pertinent squeezing parameter 
$r < 41$.
If the analysis of \cite{Hertzberg:2021rbl}  applied intact to our \emph{multi-mode-squeezed} graviton case, then from \eqref{GRsqop}, \eqref{GRsqr} one could constrain the axionic-cloud life time, and thus falsify models with too long lifetimes~\cite{porto_scales}. 

Finally, we note that, observationally, one does not have access to all of the directions of emission from the ALPs cloud. Hence, the observers need to trace out the modes (directions) one does not have  access to. Since the multi-mode squeezed graviton state has inherent \textit{directional entanglement}, the tracing out process is a non-trivial task. As discussed in \cite{Dorlis:2025amf,Mavromatos:2025ofn}, the tracing out can produce thermal states, and thus enhanced thermal noise in an interferometer or even more complex mixed states. The measurement of the quantum features of GWs constitutes a challenging step, 
because the latter
appear as noise in the interferometers and not as pure signals~\cite{Parikh:2020kfh}. The fraction of accessible modes and entanglement determine the sub-Poissonian signature. Degradation of the signature may be partially mitigated by noise-correlation measurements between spatially separated interferometers by accessing complementary directional modes \cite{Parikh:2023zat}. A  thermal state and a  traced out squeezed state have different  higher order cumulants. Measuring the third or fourth order moments of the strain noise could in principle distinguish residual quantum coherence from a purely thermal background.  A relevant approach is the one suggested in~\cite{Pang:2018eec,Abrahao:2023lle}, dealing with GW quantum noise.  The significant enhancement  of the squeezed graviton number in our black-hole superradiant case
is encouraging for further studies. 

 We conclude by elaborating briefly on the proposal of \cite{singlegrav}, which has challenged Dyson's objection~\cite{Dyson:2013hbl}, claiming the possibility of single-graviton detection through the use of modern quantum-technology sensors. 
These authors argue that the use of small, high-quality acoustic resonators, cooled down to extremely low temperatures close to their (quantum) ground state, could take advantage of the extreme weakness of gravity to 
detect, through appropriate ``quantum jumps'' in the apparatus, the exchange of a single graviton between matter in the detector and the GW, in a similar way to the photoelectric effect.
However, doubts have been expressed~\cite{Carney:2023nzz,Carney:2024dsj} on the distinguishability of the single-graviton effects from those of a classical GW: by simply counting events (``clicks'') in a detector one cannot provide unambiguous experimental proof for the existence of quantum gravitons. The quantum signature of squeezed GWs lies on the sub-Poissonian statistics for the number $N$ of such events, $\langle N^2\rangle - \langle N \rangle^2  < \langle N \rangle $, but the extreme weakness of gravity may suppress any experimental evidence for it in the case of squeezed single graviton modes. Measuring noise correlations between different detectors~\cite{Parikh:2023zat} might be a way forward. There are also proposals for quantum graviton detection~\cite{Manikandan:2025hlz} involving the use of resonant bar detectors in coincidence with GW interferometers, with a claimed capability of distinguishing fields with significant components of quantum squeezing. Our astrophysical source of such graviton states may 
significantly improve the chances of such detections,
as it amplifies QG effects. Our approach also provides a test of the hypothesis of GR as an EFT.

\vspace{-0.4cm} 
\begin{tcolorbox}[
colback=white,
colframe=black,
title=\textbf{\cy Key Prediction \cbl},
fonttitle=\bfseries,
boxrule=0.8pt,
arc=2mm,
left=2mm,
right=2mm,
top=1mm,
bottom=1mm
]

Superradiant axion clouds around rotating BH generate 
multimode squeezed graviton states containing up to 
$\langle N_{\rm gr}\rangle \sim 10^{6} - 10^{7}$ correlated gravitons. 
Such states possess characteristic polarization correlations and 
quantum-noise signatures.
Future GW interferometers may therefore detect 
non-classical features of gravitational radiation through deviations 
from classical noise statistics. Detection of these correlations would 
constitute direct observational evidence that GW are 
quantum states of the gravitational field.
 
\end{tcolorbox}

\newpage

\bibliography{QGentangl.bib}

\end{document}